\documentclass[sigconf,memoir,nonacm]{acmart}

\usepackage{orcidlink}

\setcopyright{acmcopyright}
\copyrightyear{2023}
\acmYear{2023}

\makeatletter
\renewcommand\@formatdoi[1]{\ignorespaces}
\def\@isbn{\ignorespaces}
\makeatother

\acmConference[MAPS ’23]{the 7th Annual Symposium on Machine Programming}{December 3, 2023}{San Francisco, USA}

\begin{document}

\title[LLMs Should Ask Clarifying Questions to Increase Confidence in Generated Code]{Large Language Models Should Ask Clarifying Questions to Increase Confidence in Generated Code}
\titlenote{This paper is accepted and presented at the 7th Annual Symposium on Machine Programming (MAPS '23), held on December 3, 2023 in San Francisco, CA, USA.}



\author{Jie JW Wu\,\orcidlink{0000-0002-7895-2023}}
\affiliation{
\institution{George Washington University} 
\city{}
\country{}
}
\email{jiewu@gwu.edu}

\renewcommand{\shortauthors}{Wu}

\begin{abstract}
Large language models (LLMs) have significantly improved the ability to perform tasks in the field of code generation. However, there is still a gap between LLMs being capable coders and being top-tier software engineers. Based on the observation that top-level software engineers often ask clarifying questions to reduce ambiguity in both requirements and coding solutions, I argue that the same should be applied to LLMs for code generation tasks. By asking probing questions in various topics before generating the final code, the challenges of programming with LLMs, such as unclear intent specification, lack of computational thinking, and undesired code quality, may be alleviated. This, in turn, increases confidence in the generated code. In this work, I explore how to leverage better communication skills to achieve greater confidence in generated code. I propose a communication-centered process that uses an LLM-generated communicator to identify issues with high ambiguity or low confidence in problem descriptions and generated code. I then ask clarifying questions to obtain responses from users for refining the code.

\end{abstract}




\maketitle

\begin{quote}
“Asking a good question can be valuable in and of itself, irrespective of the answer. It communicates
your respect for the other person.” \\ 
\phantom{||} - Adapted from the Iowa Peace Institute Message\\
\end{quote}

\section{Introduction}
Large language models (LLMs)~\cite{vaswani2017, svyatkovskiy2020, wang2021, feng2020}, such as OpenAI's Codex~\cite{chen2021evaluating}, AlphaCode~\cite{li2022competition}, and CodeGen~\cite{nijkamp2022codegen}, possess significant capabilities to generate code snippets from natural language requirements. However, there are several reported issues with LLMs, including problems with intent specification, problem decomposition~\cite{sarkar2022like}, code quality, and overconfidence~\cite{liu2023no,liu2023refining}, as well as usability~\cite{liang2023understanding}. These issues indicate that there is still a substantial gap between an LLM as a programming assistant \cite{rabinovich2017, ye2020, alon2019, bui2021, tufano2020} and a software engineer. As the responsibility of software developers encompasses more than just writing code, current LLMs cannot fully replace professional software developers~\cite{sarkar2022like,borji2023categorical}. At a high level, the gap lies in several critical aspects of software development beyond coding, such as effective communications, requirements, design, domain knowledge, and the broader context of relevant projects and components, etc \cite{nguyen2022, sobania2022, vaithilingam2022, siddiq2022}. In this paper, I am interested in applying the communication lens to inspect the gap, given that effective communication is a critical capability that connects all of the above-mentioned parts to coding. I study the following research question: \textit{Does asking clarifying questions increase confidence in ChatGPT-generated code?}

With this question, let us take a step back to compare the communications of LLMs and software developers. In the literature, the level of communication skills is rarely emphasized or evaluated in the field of code generation. The current LLMs are evaluated by generating code in one or multiple attempts from one-off problem descriptions, without further conversational inputs~\cite{chen2021evaluating,austin2021program,li2022competition}. This means when the input problem description is error-prone or incomplete without full context, the model has to generate the code without the chance to clarify questions that are necessary to ensure the quality and correctness of the code. On the contrary, given a software engineering task in real-world enterprises, professional developers use various ways of communication, such as asking more questions in 1:1 conversations, group meetings, and Slack channels to obtain more information and reduce ambiguity about the detailed requirements, context of the projects, and the design alternatives. Proactive and effective communication is a critical skill in practice for top-level software developers to accomplish their software engineering tasks reliably with high quality~\cite{whitehead2007collaboration,pressman2005software,mistrik2010collaborative,mcchesney2004communication,jazayeri2004education}.

Inspired by this behavior, in this work, I would like to study the potential of LLMs from the dimension of effective communication skills. As the first step toward this objective, I explore a process centered around promoting effective communication to ask more clarifying questions to refine the final generated code from LLMs. The initial exploration via an empirical example indicates that the communication-centered process is promising in boosting the communication skills of LLMs for code generation tasks. Finally, I provide conclusions and future work.

\begin{figure*}[h]
  \centering
  \includegraphics[width=0.8\textwidth]{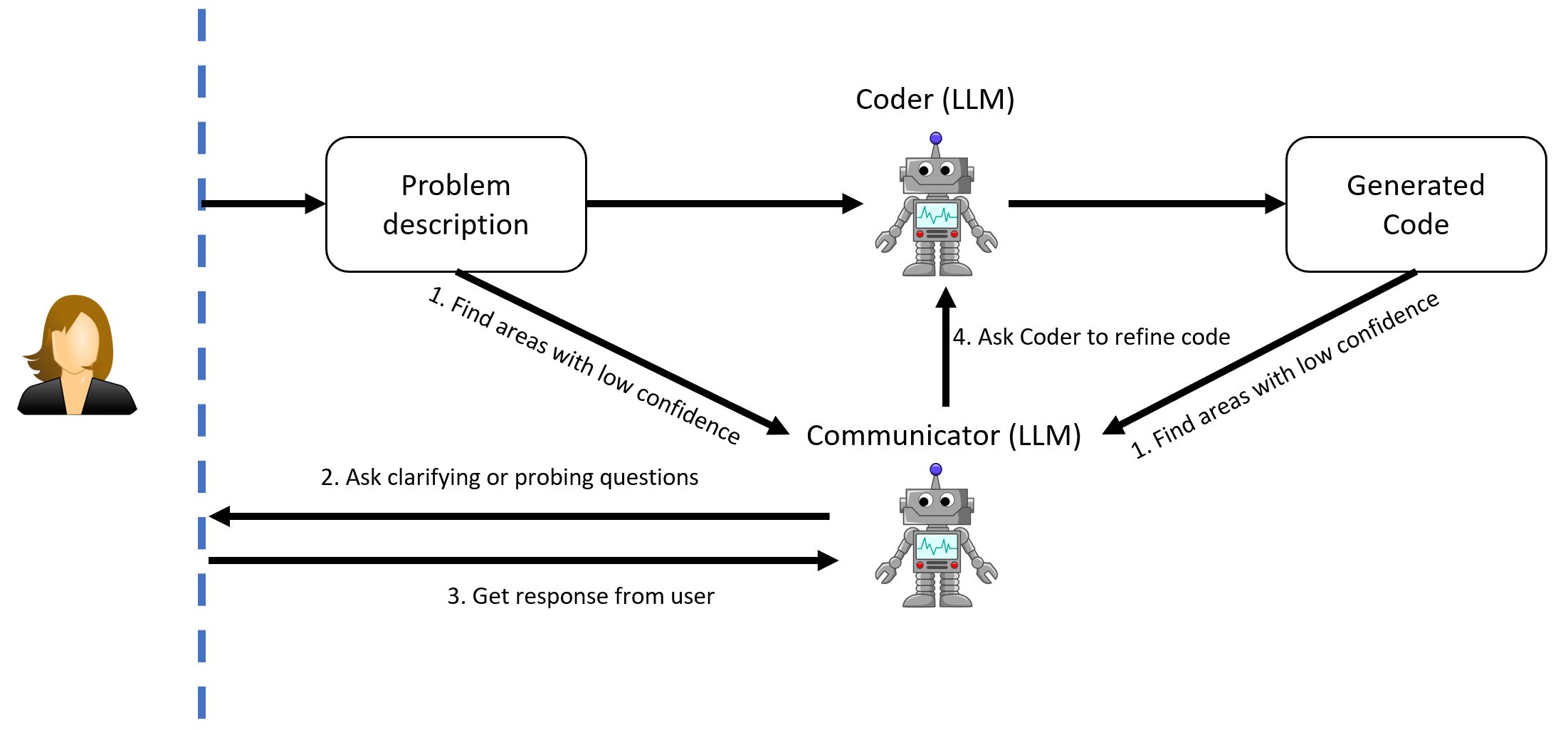}
  \caption{The visual illustration of the communication-centered process: the coder generates code according to the problem description and the communicator's information. The communicator detects the parts with low confidence where communication may help to get more information from the user. Then the communicator asks clarifying questions and gets the response. Finally, the response is sent back to the coder to refine the generated code. } 
  \label{fig:overview}
\end{figure*}

\section{Communication-Centered Process}

In this section, I describe the proposed communication-centered process for code generation tasks. The communication-centered process ultilizes two components - a \textit{coder} and a \textit{communicator}. The visual illustration of the process is shown in Figure~\ref{fig:overview}. The coder takes the problem description and, optionally, the information provided by the communicator to generate the code. The communicator review the problem descriptions and the code generated from previous iteration, then ask questions on whether certain parts are identified as low confidence and need to be clarified or probed via questions. Finally, the communicator sends back the responses from users to the coder to refine the generated code. This process is repeated until some condition is met. Each of these two components can be implemented using a LLM. 

Inspired by how the high-quality code is produced by top-level software developers, in the LLM-based code generation, my main idea for the communicator is that it should find the parts from the code and description that are typically of higher uncertainty and low confidence. These parts can be improved by getting more useful information from users via good clarifying and probing questions. The communicator asks questions focused on various aspects of software development concerning the high standards of final output code, such as intent specification, intent clarification, requirement disambiguity, coding style etc. I use an example in the next section to illustrate how the coder works with the communicator in detail.

\section{Example Usage}
\begin{figure*}[h]
  \centering
  \includegraphics[width=\textwidth]{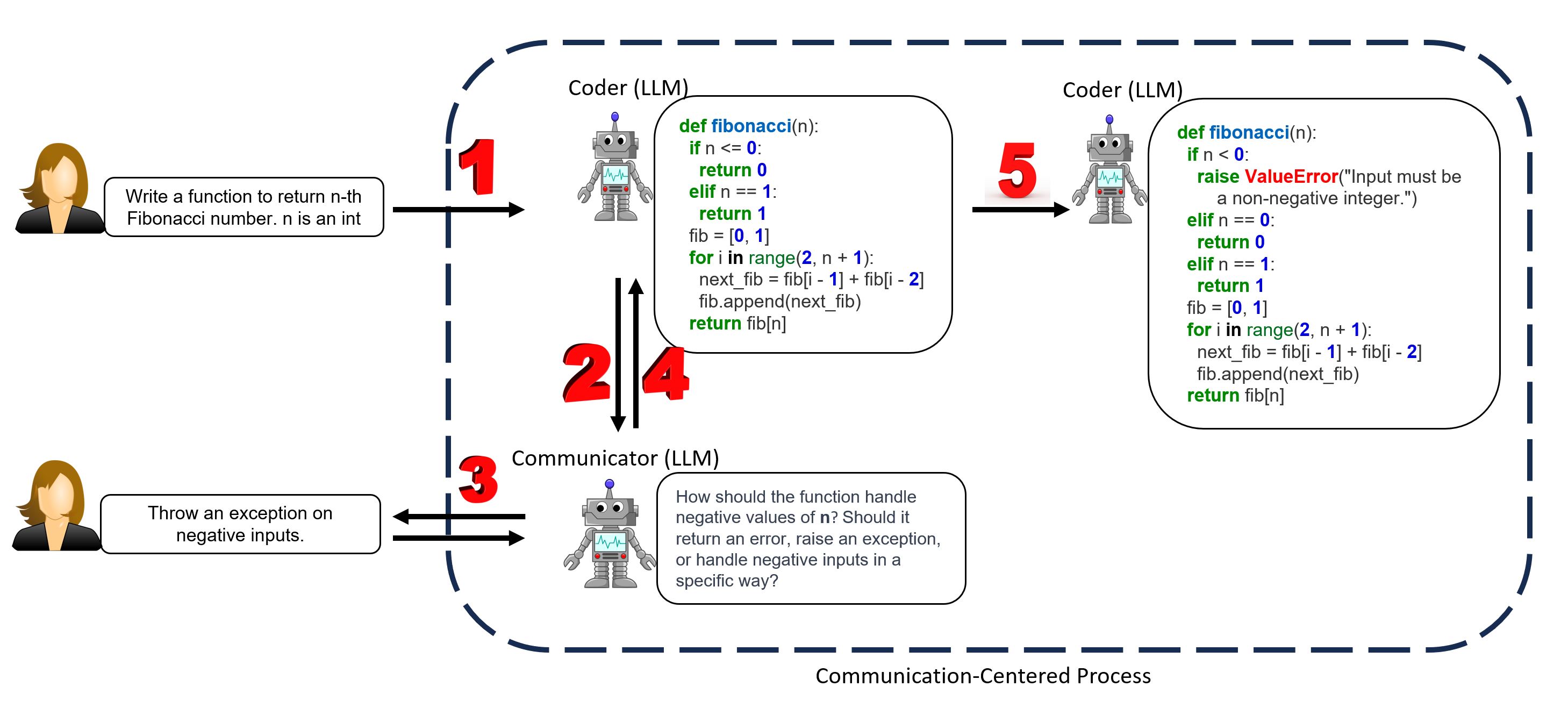}
  \caption{An example of the communication-centered process. }
  \label{fig:example1}
\end{figure*}

I showcase a practical example using the communication-centered process for code generation tasks. In this example, I use ChatGPT 3.5 as the LLM for the coder and the communicator. As shown in Figure~\ref{fig:example1}, a user is looking for a code snippet to return n-th Fibonacci number. Within the process, the coder LLM first generated an initial code snippet. From the initial code output that, although ChatGPT 3.5 as the coder generated the code as required, several topics on this can be further elaborated and discussed. In this case, the communicator analyzed the problem description and the initial code to generate a list of clarifying questions. The top question is concerned with the negative input, which was not mentioned in the problem description. The user responded the question to set the expectation on the case of the negative input. Then, the coder refined the code with that additional information. I only use one iteration in the example, but in theory this loop can continue until some criteria is met.                 

The key part of the process is how the communicator interacts with the user and the coder. For the communicator, I use ChatGPT 3.5 as the LLM, which is same as the coder. The communicator uses zero-shot prompting to generate the clarifying questions from the problem description and the code from the previous iteration. The example of prompt from Figure~\ref{fig:example1} is described below. 
\\ \\
\fbox{\begin{minipage}{25em}
 You are an expert in software engineering. You will be given the problem description and current code of a coding task. You will generate a list of clarifying questions that may result in refining the code. \\ \\
\#\#\# Problem Description\\
"write a function to return n-th Fibonacci number. n is an int "\\ \\
\#\#\# Generated Code From Previous Iteration \\ 
def fibonacci(n):\\
\phantom{||}if n <= 0:\\
\phantom{||||}return 0\\
\phantom{||}elif n == 1:\\
\phantom{||||}return 1\\
\phantom{||}fib = [0, 1]\\
\phantom{||}for i in range(2, n + 1):\\
\phantom{||||}next\_fib = fib[i - 1] + fib[i - 2]\\
\phantom{||||}fib.append(next\_fib)\\
\phantom{||}return fib[n]\\
\end{minipage}}
\\ \\
\\ \\
The output list of clarifying questions for the communicator in the above example is shown in Table~\ref{tab:clarifying_questions}. With the simple zero-shot prompt, the communicator is already able to automatically cover a number of topics that deserve asking questions. As the future extension, one could modify the prompt to focus on topics that are more helpful for their customized usage or for the quality of generated code. For example, topics such as code context, coding style, non-functional requirements, code analysis, explanation are mentioned as the current weakness of the code output of LLMs~\cite{liang2023understanding}. So, these topics can be better prioritized in the prompt of the communicator. From this mere example, I see big potential that boosting communication skills can lead to improvements of the final code output in several topics such as negative input, documentation, etc.  

\begin{table*}[h]
\centering
\renewcommand\arraystretch{1.5} 
\begin{tabular}{|c|p{10cm}|}
\hline
\textbf{Topic} & \textbf{Clarifying Questions Generated by the Communicator LLM} \\
\hline
Input Validation &
Should input validation be part of the function? \\
\cline{2-2}
&
If so, what should be done if n is not an integer? \\
\hline
Error Handling &
How should errors be handled if n is not a valid integer or if it's negative? \\
\cline{2-2}
&
Should the function raise exceptions, return error codes, or follow a different error-handling strategy? \\
\hline
Performance Requirements &
Are there any performance requirements or constraints for this function? \\
\cline{2-2}
&
Is there a need for optimizing the function for efficiency, especially for large values of n? \\
\hline
Function Signature &
Are there any specific naming conventions or requirements for the function name and parameters? \\
\cline{2-2}
&
Should the function take any additional parameters, options, or configuration settings? \\
\hline
Output Type &
Is there a preferred return type for the function, such as an integer, a string, or another data type? \\
\hline
Default Values &
Should the function provide default values for n if it's not provided by the caller? \\
\hline
Documentation &
What documentation standards should be followed for this function, such as docstrings, comments, or external documentation? \\
\hline
Testing &
Are there specific test cases or test scenarios that should be considered when testing the function? \\
\cline{2-2}
&
Should the function be tested for edge cases, such as very large values of n? \\
\hline
Language and Environment &
Is there a specific programming language or environment in which this function should be implemented? \\
\cline{2-2}
&
Are there any platform or compatibility requirements? \\
\hline
Algorithm Choice &
Is there a preference for using a specific algorithm for calculating Fibonacci numbers, such as an iterative or recursive approach? \\
\cline{2-2}
&
Should the function allow for different algorithm choices or optimizations? \\
\hline
Usability and Extensibility &
Should the function be designed for easy integration into other codebases? \\
\cline{2-2}
&
Are there considerations for future extensions or modifications to the function? \\
\hline
Security Considerations &
Are there any security considerations related to the input or output of the function that need to be addressed? \\
\hline
Concurrency and Parallelism &
Does the function need to support concurrent or parallel execution for improved performance? \\
\hline
Version Control and Deployment &
How will the function be versioned, deployed, and maintained in the software development lifecycle? \\
\hline
Use Case and Context &
What is the primary use case or context in which this function will be used? \\
\cline{2-2}
&
Understanding the broader context can help in designing the function to meet specific needs. \\
\hline
\end{tabular}
\caption{Example of clarifying questions in the Communicator LLM from Figure~\ref{fig:example1}.}
\label{tab:clarifying_questions}
\end{table*}

\section{Background and Related Work}
\textbf{Code Generation with Large Language Models.} In recent years, the field of code generation has seen a significant shift with the large language models. For example, Codex~\cite{chen2021evaluating}, fine-tuned on GPT-3~\cite{brown2020language} on a large corpus of source code data, is capable of generating code for 47/164 problems in the HumanEval dataset in single run, a benchmark for code generation task. Codex became the core model for the Copilot~\cite{ziegler2022productivity}, an AI-powered coding assistant developed by GitHub. After Codex, a couple of models similar to Codex but with smaller size were then developed, including GPT-J~\cite{wang2021gpt}, CodeParrot~\cite{huggingface-codeparrot}, PolyCoder~\cite{xu2022systematic}. AlphaCode~\cite{li2022competition}, with size comparable to Codex, was trained on Github data and fine-tuned on competition-level programing problems. It exceeded half of the competitors in coding competitions of CodeForces, a well-known online competitive programming platform. CodeGen~\cite{nijkamp2022codegen} was trained on both natural language and programming language data for code generations with multi-turn prompts. However, the level of communication skills of these models are not emphasized and evaluated. These models are evaluated by generating code in one or multiple attempts from one-off problem descriptions, without further information from conversations. Therefore, when the input problem description is error-prone or incomplete, the model still has to generate the code without the chance to clarify critical questions. My work serves as an exploration to address this usability problem.

\textbf{Self-Correct Large Language Models.} Recently, a promising approach to improve the output efficiency of large language models is \textit{self-correction}~\cite{pan2023automatically}. In self-correction approach, the LLM uses the feedback guided or prompted by itself to refine their results. One popular category of work uses human feedback to refine their results directly~\cite{kreutzer2018, glaese2022, ouyang2022, scheurer2023, fernandes2023}. Other studies employed different strageties to self-correct LLMs using automated feedback such as self-training \cite{huang2022, bai2022b}, generate-then-rank \cite{he2023, weng2023}, feedback-guided decoding \cite{yang2022a, xie2023}, iterative post-hoc revision \cite{zhang2023a, jiang2023}, etc. My work in this paper is also under the category of self-correction using both human feedback and automated feedback, but with a new perspective of improving communication skills for code generation tasks.

\section{Conclusion and Future Work}
In this exploratory paper, I showed an initial step on the problem of how to increase the communication skills of LLMs to elevate the final generated code via clarifying and probing questions. I argue that the proficiency of communication skills of LLMs is necessary for generating code with high standards. Elevated communication skills should be viewed as an important factor toward bridging the gap between LLMs and top-notch software developers. Specifically, this effort will not only increase confidence of generated code, but also gain the trust of users using them as a programmin assistant. Although it needs additional efforts of conversational input from users, I believe it is still necessary and worthwhile. As a first step toward this effort, I explored a communication-centered process that ultilizes a coder LLM and a communicator LLM that work together to ask clarifying questions and refine the generated code. By generating clarifying questions in prompts via the communicator, there is hope to raise the capability of communication skills to produce code with higher confidence. However, several challenges still exist along this line of research and future work is needed, as described below.

\textbf{Evaluation benchmark of communication skills for LLM-based code generation.} This is needed to objectively quantify the capability of communication skills of LLM on code generation tasks and software engineering tasks. As aforementioned, the existing evaluation work of LLMs for code generation mainly focus on solving algorithm problems without additional converstational input. In my case, as future work, I will target at benchmark that the reveal how effective is the communcation ability of the model. For example, creating a dataset with blurred and noisy problem descriptions that hide some critical information is an interesting direction. In this setting, the model should ask the right questions rather than directly generate low-quality code. 

\textbf{Improving communication skills of LLMs.} Besides benchmark, techniques to further improve the communication skills of LLMs can be the next steps as future work. Since the prompts are critical to the success of LLMs, one of the important future work is to design effective instructions w.r.t. communication skills. Another interesting angle is to study how to tune the model to switch between under-communicating, effective-communcating and over-communicating. I envision that different AI programming agents in future will have various levels and styles of communication ability. This work can be seen as the first step toward improving communication skills of LLMs for code. 

\section{Acknowledgements}
I would like to thank the anonymous reviewers for their constructive feedback and suggestions. I also thank Vincent Hellendoorn for his insightful feedback after presenting this work at MAPS 2023. 

\bibliographystyle{ACM-Reference-Format}
\bibliography{sample-base}


\end{document}